\documentstyle[12pt]{article}

\def\hybrid{\topmargin -20pt \oddsidemargin 0pt
\headheight 0pt \headsep 0pt
\textwidth 6.25in 
\textheight 9.5in 
\marginparwidth .875in
\parskip 5pt plus 1pt \jot = 1.5ex}

\hybrid

\def\baselinestretch{1.2}

\catcode`\@=11

\def\marginnote#1{}
\newcount\hour
\newcount\minute
\newtoks\amorpm
\hour=\time\divide\hour by60
\minute=\time{\multiply\hour by60 \global\advance\minute by-\hour}
\edef\standardtime{{\ifnum\hour<12 \global\amorpm={am}%
\else\global\amorpm={pm}\advance\hour by-12 \fi
\ifnum\hour=0 \hour=12 \fi
\number\hour:\ifnum\minute<10 0\fi\number\minute\the\amorpm}}
\edef\militarytime{\number\hour:\ifnum\minute<10 0\fi\number\minute}

\def\draftlabel#1{{\@bsphack\if@filesw {\let\thepage\relax
\xdef\@gtempa{\write\@auxout{\string
\newlabel{#1}{{\@currentlabel}{\thepage}}}}}\@gtempa
\if@nobreak \ifvmode\nobreak\fi\fi\fi\@esphack}
\gdef\@eqnlabel{#1}}
\def\@eqnlabel{}
\def\@vacuum{}
\def\draftmarginnote#1{\marginpar{\raggedright\scriptsize\tt#1}}

\def\draft{\oddsidemargin -.5truein
\def\@oddfoot{\sl preliminary draft \hfil
\rm\thepage\hfil\sl\today\quad\militarytime}
\let\@evenfoot\@oddfoot \overfullrule 3pt
\let\label=\draftlabel
\let\marginnote=\draftmarginnote
\def\@eqnnum{(\theequation)\rlap{\kern\marginparsep\tt\@eqnlabel}%
\global\let\@eqnlabel\@vacuum} }

\def\preprint{\twocolumn\sloppy\flushbottom\parindent 2em
\leftmargini 2em\leftmarginv .5em\leftmarginvi .5em
\oddsidemargin -.5in \evensidemargin -.5in
\columnsep .4in \footheight 0pt
\textwidth 10.in \topmargin -.4in
\headheight 12pt \topskip .4in
\textheight 6.9in \footskip 0pt
\def\@oddhead{\thepage\hfil\addtocounter{page}{1}\thepage}
\let\@evenhead\@oddhead \def\@oddfoot{} \def\@evenfoot{} }

\def\numberbysection{\@addtoreset{equation}{section}
\def\theequation{\thesection.\arabic{equation}}}

\def\underline#1{\relax\ifmmode\@@underline#1\else
$\@@underline{\hbox{#1}}$\relax\fi}

\def\titlepage{\@restonecolfalse\if@twocolumn\@restonecoltrue\onecolumn
\else \newpage \fi \thispagestyle{empty}\c@page\z@
\def\thefootnote{\fnsymbol{footnote}} }

\def\endtitlepage{\if@restonecol\twocolumn \else \newpage \fi
\def\thefootnote{\arabic{footnote}}
\setcounter{footnote}{0}} 

\catcode`@=12
\relax

\def\figcap{\section*{Figure Captions\markboth
{FIGURECAPTIONS}{FIGURECAPTIONS}}\list
{Figure \arabic{enumi}:\hfill}{\settowidth\labelwidth{Figure
999:}
\leftmargin\labelwidth
\advance\leftmargin\labelsep\usecounter{enumi}}}
 \relax
\def\tablecap{\section*{Table Captions\markboth
{TABLECAPTIONS}{TABLECAPTIONS}}\list
{Table \arabic{enumi}:\hfill}{\settowidth\labelwidth{Table
999:}
\leftmargin\labelwidth
\advance\leftmargin\labelsep\usecounter{enumi}}}
 \relax
\def\reflist{\section*{References\markboth
{REFLIST}{REFLIST}}\list
{[\arabic{enumi}]\hfill}{\settowidth\labelwidth{[999]}
\leftmargin\labelwidth
\advance\leftmargin\labelsep\usecounter{enumi}}}
 \relax
\makeatletter
\newcounter{pubctr}
\def\publist{\@ifnextchar[{\@publist}{\@@publist}}
\def\@publist[#1]{\list
{[\arabic{pubctr}]\hfill}{\settowidth\labelwidth{[999]}
\leftmargin\labelwidth
\advance\leftmargin\labelsep
\@nmbrlisttrue\def\@listctr{pubctr}
\setcounter{pubctr}{#1}\addtocounter{pubctr}{-1}}}
\def\@@publist{\list
{[\arabic{pubctr}]\hfill}{\settowidth\labelwidth{[999]}
\leftmargin\labelwidth
\advance\leftmargin\labelsep
\@nmbrlisttrue\def\@listctr{pubctr}}}
 \relax
\makeatother
\newskip\humongous \humongous=0pt plus 1000pt minus 1000pt

\newif\ifdtup

\relax

\def\be{\begin{equation}}
\def\ee{\end{equation}}
\def\ba{\begin{eqnarray}}
\def\ea{\end{eqnarray}}

\def\del{\partial}

\def\r{\rho}

\def\m{\mu}
\def\n{\nu}

\def\L{\Lambda}

\def\no{\noindent}

\def\IR{\relax{\rm I\kern-.18em R}}

\def\IR{\relax{\rm I\kern-.18em R}}
\def\inv{^{\raise.15ex\hbox{${\scriptscriptstyle -}$}\kern-.05em 1}}

\def\tL{{\tilde L}}

\begin{document}

\renewcommand{\theequation}{\arabic{equation}}

\newcommand{\beq}{\begin{equation}}
\newcommand{\eeq}[1]{\label{#1}\end{equation}}
\newcommand{\ber}{\begin{eqnarray}}
\newcommand{\eer}[1]{\label{#1}\end{eqnarray}}
\newcommand{\eqn}[1]{(\ref{#1})}
\begin{titlepage}
\begin{center}

\hfill hep--th/0003281\\

\vskip .9in

{\large \bf Hyperbolic Spaces in String and M-Theory}

\vskip 0.4in

{\bf A. Kehagias$^1$}
\phantom{x} and\phantom{x} {\bf J.G. Russo$^2$}
\vskip 0.1in
{\em ${}^1$Physics Dept. National Technical University \\
GR-157 73 Zografou Campus, Athens, Greece\\
${}^2$ Departamento de F\'\i sica, Universidad de Buenos Aires\\
Ciudad Universitaria, Pabell\'on I, 1428 Buenos Aires
\vskip 0.1in
{\tt kehagias@mail.cern.ch,~ russo@df.uba.ar
}}\\
\vskip .1in

\end{center}

\vskip .4in

\centerline{\bf Abstract }

\no

We describe string-theory and $d=11$ supergravity solutions
involving symmetric spaces  of constant negative
curvature.
Many examples of
non-supersymmetric
string compactifications on
hyperbolic spaces $H_r$ of finite volume are given in terms of
suitable cosets of the form $H_r/\Gamma $, where $\Gamma $ is a
discrete group. We describe in some detail the cases of
the non-compact hyperbolic spaces $F_2$ and $F_3$, representing the
fundamental regions of $H_2$  and $H_3$ under $SL(2,Z)$ and the
Picard group, respectively. 
By
writing $AdS$ as a  $U(1)$ fibration, we obtain new solutions
where $AdS_{2p+1}$ gets untwisted by T-duality to ${\bf R}\times
SU(p,1)/(SU(p)\times U(1))$. Solutions with time-dependent dilaton
field are also constructed by starting with a solution with
NS5-brane flux over $H_3$. 
A new class of  non-supersymmetric conformal field theories
can be defined via holography.

\no

\vskip 2.5cm 
March 2000\\
\end{titlepage}
\vfill
\eject

\def\baselinestretch{1.2}
\baselineskip 16 pt
\noindent

\def\tT{{\tilde T}}
\def\tg{{\tilde g}}
\def\tL{{\tilde L}}

\def\del{\partial }


\section{Introduction}

Maximally symmetric spaces play an important role in supergravity \cite{DNP}
and in superstring theory \cite{Mald}, and they  naturally arise as the
near-horizon region
of black brane geometries \cite{horo,GT}.
Maximally symmetric spaces
(with no more than one time dimension) are the de Sitter $dS_p$
and anti de Sitter $AdS_p$ spaces, $n$-spheres $S^n$ and
$r$-dimensional hyperboloids $H_r$. Although supergravity
solutions describing spheres and $AdS$ spaces have been
extensively investigated \cite{DNP}, not much has been done for solutions
involving the de Sitter
and hyperbolic geometries.
There are several obvious reasons for this.  De Sitter spaces always break
supersymmetry, and they do not
describe universes with zero cosmological constant.
Hyperbolic spaces, such as the upper half
$r$-plane, admit Killing spinors \cite{BF,FY,LP} but they have infinite
volume.
Therefore, they do not seem useful for describing internal spaces
in string compactifications.
Finite volume examples can be constructed by cosets of the form
$H_r/\Gamma$, where $\Gamma $ is a discrete subgroup of the
isometry group of $H_r$ \cite{M,ARVO,kaloper}.
However, dividing by $\Gamma$
breaks all supersymmetries (see sect.~3).

We shall show that there is a
large class of new string compactifications involving $H_r$ or cosets
$H_r/\Gamma $,
which are  regular (away possible orbifold points), where $\alpha '$ string
corrections  can be made very small for a sufficiently large radius of
$H_r$.
Among the possible applications of these solutions, we  explore: a) the
definition  of new ${\cal N}=0$ conformal field theories in four
dimensions via holography (see also \cite{kach,NV}); b) obtaining new models
for dimensional
reduction in non-compact spaces.

This paper is organized as follows. In sect.~2, we describe
solutions of eleven-dimensional supergravity representing spaces
which are direct products containing $H_r$ factors and discuss the
supersymmetry. Analogous solutions in type II supergravity are also given.
In sect.~3, we discuss  finite-volume compactifications. 
In sect.~4 we compute the Newton gravitational potential for
spaces
having the hyperbolic 3-manifold $H_3$ as internal space.
 Because the Kaluza-Klein eigenvalue spectrum on $H_3$ is continuous
with a mass gap, the Newton potential
has a Yukawa-type behavior at large distances.
Hyperbolic spaces naturally appear in theories with other spacetime signatures
\cite{cremmer,hull,hullk}
(e.g. by Wick rotating  $AdS$ solutions to Euclidean space) or 
in solutions with imaginary gauge fields, which can be
interpreted as solutions with real gauge fields in new type II* theories \cite{hull},
where the signs of  kinetic terms of RR gauge fields are reversed.
In sect.~5, we consider two different
approaches for obtaining new solutions in these theories.
One of them consists in writing $AdS_{2p+1}$ as a $U(1)$ fibration and performing 
T-duality in a time-like direction.
In sect.~6, we discuss the dual conformal field
theories  in  compactifications that are direct product of
anti de Sitter,
hyperbolic manifolds of finite volume and spheres.
Finally, we calculate in the appendix the Euler
number of the space $SU(2,1)/SU(2)\times U(1)$.

\section{Hyperbolic manifolds in Supergravity}
\subsection{Eleven-dimensional Supergravity}

The eleven-dimensional graviton multiplet contains the graviton $g_{MN}$,
the antisymmetric three-form $A_{MNK}$ and the gravitino $\Psi_M$
($M,N,...=0,...,10$). The bosonic part
of the eleven-dimensional supergravity Lagrangian is
\be
{\cal L}=\frac{1}{2\kappa_{11}^2}\sqrt{g}
\left(R-{1\over 2\cdot 4!}F_{MNPQ}F^{MNPQ}\right)
-{1\over  {12}^4}\epsilon^{M_1\ldots M_{11}} A_{M_1M_2M_3}
F_{M_4\ldots M_7}F_{M_8\ldots M_{11}}
\, . \label{action}
\ee
A well-known solution to the equations of motions
\ba
R_{MN}&=&\frac{1}{12}\left(F_{MPQR}{F_N}^{PQR}-\frac{1}{12}g_{MN}F^2\right)\
,
\label{Eins} \\
\nabla_M F^{MNPQ}&=&-{1\over 1152}\epsilon^{NPQR_1...R_8}F_{R_1...R_4}
F_{R_5...R_8}\, , \label{F}
\ea
is provided by the
Freund-Rubin ansatz for the antisymmetric field strength
\ba
F_{mnpq}&=&6m_0\,\epsilon_{mnpq}~~~ \mbox{for}~~~ m,n,...=7,...,10\,
,\nonumber \\
F_{MNPQ}&=&0\, , ~~~~\mbox{otherwise.} \label{FR}
\ea
By substituting this ansatz into the field equations (\ref{Eins})
we get
\ba
R_{\m\n}&=&-6m^2_0 g_{\m\n}\, , ~~~~~ \m,\n=0,...6\, , \nonumber \\
R_{mn}&=&12m^2_0 g_{mn}\, . \label{sol}
\ea
The requirement of unbroken supersymmetry, i.e., the vanishing
of the gravitino  transformation
\be
\delta\Psi_M=\nabla_M\epsilon-{1\over 288} \left({\Gamma_M}^{PQRS}-8
{\delta_M}^P\Gamma^{QRS}\right)\epsilon F_{PQRS}\, ,
\ee
 for the ansatz (\ref{FR}), is equivalent to the existence of
 $SO(1,6)$ and $ SO(4)$ Killing spinors $\theta$ and $\eta$, respectively,
 which satisfy
\ba
\nabla_\m \theta &=& \pm {1\over 2} m_0 \gamma_\m \theta\, , \label{sf} \\
\nabla_m \eta &=&  \pm m_0 \gamma_m \eta\, , \label{shifts}
\ea
where $\gamma_\m~(\gamma_m)$ are
$SO(1,6)~(SO(4))$ $\gamma$-matrices. Thus, although
eq.~(\ref{sol}) is solved for
 $ Y^7 \times X^4$ where $Y^7$ and $X^4$ are Einstein spaces of negative and
positive  curvature, respectively, only  those spaces
that admit  Killing spinors obeying eqs.~(\ref{sf}),~(\ref{shifts}) are
supersymmetric. The integrability condition of
eqs.(\ref{sf}),~(\ref{shifts}) are
\be
W_{\mu\nu\rho\sigma}\gamma^{\rho\sigma}\theta=0\, ,
~~~~W_{mnpq}\gamma^{pq}\eta=0\, , \label{W} \ee where
$W_{\m\n\rho\sigma}, ~W_{mnpq}$ are  the Weyl tensors of
$Y^7,~X^4$, respectively. Thus, obvious supersymmetric
 examples for $X^4$ include the round
four-sphere $S^4$ and its orbifolds $S^4/\Gamma$, where $\Gamma$
is an appropriate discrete group \cite{FKPZ}. For the $Y^7$ space one can
take
the anti-de Sitter space $AdS_7$, which preserves supersymmetry as
well, and leads to the $AdS_7\!\times\!S^4$ vacuum of
eleven-dimensional supergravity.

 In addition to  solutions of the form $AdS_7\times X^4$,
there are other solutions to
eq.~(\ref{sol})  involving hyperbolic manifolds.
The spaces $AdS_{7-r}\!\times\!H_{r}\!\times\! S^4$, where $H_r,~r\geq 2$
is the $r$-dimensional upper-half space with metric:
\be
ds^2=dH_r^2\equiv {R_0^2\over x_0^2}\big( dx_0^2+dx_1^2+...+dx_r^2\big)
\, , \ \ \ \ \ \ \ \ R_0^2={r\over 6m^2_0}\ ,
\label{poin}
\ee
solve eqs.~(\ref{Eins}), (\ref{F}). 
Thus, the following backgrounds are all vacua of
eleven-dimensional supergravity:
\smallskip

\noindent
i) $AdS_2\!\times\! H_5\!\times\!S^4$ ;

\noindent
ii) $AdS_2\!\times\! H_2 \!\times\!H_3\!\times\!S^4$ ;

\noindent
iii) $AdS_3\!\times\! H_2\!\times\!H_2\!\times\!S^4$ ;

\noindent
iv) $AdS_3\!\times\! H_4\!\times\!S^4$ ;

\noindent
v) $AdS_4\!\times\! H_3\!\times\!S^4 $ ;

\noindent
vi) $AdS_5\!\times\! H_2 \!\times\!S^4$ .

Neither of these  backgrounds are supersymmetric. 
Indeed, they do not
satisfy eq.~(\ref{W}). [Note that the product-space of conformally
flat spaces is not conformally flat].
However,
although supersymmetry automatically guarantees the stability of
the background, its absence does not necessarily implies
instability. We recall for example the cases of
$AdS_4\times M^{pqr}$ \cite{WWW} and
$AdS_4\times Q^{pqr}$ \cite{CF} compactifications of eleven-dimensional
supergravity, which are stable for a certain range of $p,q$
\cite{DD,PP}. A general result of the stability analysis in
\cite{DD} was that, a generic $AdS_4\times M_7$
compactification, where the Einstein space is a product of an
$n$-dimensional manifold $M_{(1)}^n$ and a $7-n$-dimensional one
$M_{(2)}^{7-n}$,
$M_7=M_{(1)}^n\times M_{(2)}^{7-n}$, is unstable. This is due to the fact
that the stability of the $AdS_4$ vacuum depends on the
eigenvalues of the Lichnerowicz operator $\Delta_L$ in the
internal space $M_7$ acting on transverse, tracefree symmetric
tensors $h_{mn}$. The background is supersymmetric if $\Delta_L$
is strictly positive (in fact $\Delta_L$ should be bigger than a
multiple of $m_0^2$). However, for product spaces, $\Delta_L$ has a
zero eigenvalue with eigenfunction $h_{mn}=diag(g_{m_1n_1},
-{n\over(7-n)}
g_{m_2n_2})$, where $g_{m_1n_1},
g_{m_2n_2}$ are the metrics on $M_{(1)}^n, M_{(2)}^{7-n}$,
respectively. Clearly, this eigenfunction corresponds
to an  expanding $M_{(1)}^n$  and a contracting $M_{(2)}^{7-n}$
with constant total volume of $M_{(1)}^n\times M_{(2)}^{7-n}$.
Thus, the instability is due to volume-preserving metric
deformations of the internal space.

All the (i)-(vi) vacua are clearly of the form described above,
so it may appear that they are unstable if  $H_r$ is replaced by a finite volume coset
$H_r/\Gamma$, as it will be described in the next section.
However, this is not the case. Let us consider for example the
 case (v) involving $AdS_4$. Fixing the $AdS$ radius, the corresponding
radius of $H_r$ is also fixed and instabilities can only be produced by
metric perturbations on $S^4$ which preserve the Einstein metric. Such perturbations
do not exist and  although the
background is of product form, it is stable. Of course, a
definite statement for the stability of (i)-(vi) needs a
case-by-case study of the spectrum. However, as the latter has not
been studied yet, at least the potentially
dangerous volume-preserving metric deformations of the internal
space are absent.

Other non-supersymmetric compactifications can be obtained by replacing
$H_r$ in the spaces i) -- vi) by any Einstein space of the same negative curvature.
In particular, one has the solution 
\smallskip

\noindent vii) $AdS_3\times \Sigma_4 \times S^4$

\noindent where $\Sigma_4 $ is the coset space $SU(2,1)/\big( SU(2)\times U(1) \big) $.
This space will appear again in sect.~5.1 (see also the appendix).

\subsection{Hyperbolic manifolds in type II supergravity}


Simple type IIB supergravity backgrounds with D3 brane charge
and constant dilaton are the following ones:

\smallskip

\noindent
i) $AdS_3\times H_2\times S^5$ ;

\noindent
ii) $AdS_2\times H_3\times S^5$ .

\noindent Like $AdS_5\times S^2\times S^3$, these spaces have no Killing
spinors.
In particular, in the case ii),
the metric is given by
\be
ds^2_{10B}={R^2\over 4z^2}(-dt^2+dz^2)+{R^2\over 2v^2}(dx^2+dy^2+dv^2)+ R^2
d\Omega_5^2\ .
\ee
The 5-form self-dual field strength is the same as in the $AdS_5\times S^5$
solution.
The isometry group is $SO(1,2)\times SO(1,3)\times SO(6)$.
Although the solution is not supersymmetric, it is regular
everywhere (being a direct product of Einstein spaces),
and $\alpha '$ corrections can be made very small
for sufficiently large radius $R$.

\section{Finite volume cosets $H_r/\Gamma $}

In the previous section we have seen some supergravity solutions involving
factors $AdS_p$ and $H_r$. The space $H_r$ has infinite volume (with respect
to the Poincar\'e metric).
Thus there are no normalizable modes for any field
in $H_r$, and the bulk as well as the boundary
theories are both empty.
Non-empty bulk and boundary field theories can be obtained
by forming the coset $H_r/\Gamma$ where $\Gamma$ is a discrete subgroup
of the isometry group  of the upper-half plane. We shall now discuss
some specific examples.

\subsection{ Killing spinors}


The supergravity solutions involving the $H_r$ manifolds discussed in
sect.~2 are
not supersymmetric. However,
our analysis does not exclude that
there could be more complicated backgrounds involving $H_r$, where some
supersymmetries are unbroken.
The Poincar\' e manifolds $H_r$ admit Killing spinors (explicit
expressions
can be found in \cite{FY}), and a question of general interest is whether they
survive under
the orbifolding by the discrete group $\Gamma$. For this we may describe the
upper-half space $H_r$ as the hypersurface (taken for simplicity to
 be of constant curvature $-1$)
\be
-X_0^2+X_1^2+...+X_r^2=-1\, .
\ee
The isometry group of $H_r$ is then clearly $SO(1,r)$ and the group $\Gamma$
is in general a subgroup of $SO(1,r)$, which may or may not have  fixed
points.
There are two cases to be discussed. If $r=even$, then a Killing spinor is
in the spinorial representation of $SO(1,r-1)$. Then, if $\Gamma $ is a
subgroup of
$SO(1,r-1)$, but it is not a subgroup of   $SO(1,r-3)$,
then there
are no surviving Killing spinors. The latter exist if $\Gamma\subset
SO(1,r-3)$.
However, in this case $H_r/\Gamma$ will still be of infinite volume.
Thus there are no finite volume cosets $H_r/\Gamma $, with $r=even$,
with unbroken supersymmetries.
In particular, any supergravity compactification involving
finite volume cosets $H_2/\Gamma $ and $H_4/\Gamma $
will not preserve any supersymmetry.

The second case is when $r=odd$. In this case, there are two Killing spinors
on $H_r$ in the spinorial representation of $SO(1,r-1)$. These two Killing
spinors are also
Weyl spinors of the isometry group $SO(1,r)$, so that they form an
irreducible
Dirac spinor of $SO(1,r)$.
If $\Gamma$ is not a subgroup $ SO(1,r-1)$,  then all supersymmetries
are broken, while if $\Gamma$ is a subgroup $ SO(1,r-1)$, then half of the
supersymmetries
survive. \footnote{The $r=odd$ case may be compared to the
coset $S^{2n+1}/\Gamma$ \cite{DGM,kach,NV}. $S^5/\Gamma$ is supersymmetric
if $\Gamma\subset
SU(3),\ SU(2)$, while is not supersymmetric if $\Gamma $ is a maximal
subgroup of
$SU(4)$.}
In this case, it may seem that the space $H_r/\Gamma$ can  be of finite
volume. In the particular case of $H_3$, we will explicitly see below that
there
is no choice of $\Gamma $ that lead to a finite volume supersymmetric space.


\subsection{Solutions of the form $AdS^5\!\times\!H_2/\Gamma\!\times \!
S^4$}


Let us consider in first place the
 $AdS^5\!\times\!H_2/\Gamma\!\times \! S^4$ solution of eleven-dimensional
supergravity.
The metric and four-form field strength are given by
\be
ds^2_{11}={2R_0^2\over 3z^2}(-dt^2+dx_1^2+dx_2^2+dx_3^2+dz^2)+
{R_0^2\over 6v^2}(dx^2+dv^2)+{R_0^2\over 4}d\Omega_4^2\ ,
\ee
\be
\hat F_4=6 R_0^3 \Omega_4\ ,
\ee
where $d\Omega_4^2$ is the metric on $S^4$ and we have taken the Poincar\'e
metric for the upper half-plane $H_2$ ($v>0$).
A finite volume  hyperbolic space can be obtained
by forming the coset $H_2/\Gamma$, where $\Gamma$ is a discrete subgroup
of the isometry group $SL(2,{\bf R})$ of the upper-half plane. Thus,
$\Gamma$ must be a Fuchsian group, and the requirement of finiteness of the
volume of $H_2/\Gamma$ is equivalent to the Fuchsian group $\Gamma$ to be of
the first kind. Examples are compact hyperbolic manifolds
representing genus 2 Riemann surfaces.

A non-compact space of finite volume can be constructed by taking
$\Gamma =SL(2,{\bf Z})$, so that $H_2/\Gamma$ is the fundamental domain
$F_2$ of $SL(2,{\bf Z})$.
This case provides a rather simple setup for dimensional reduction,
since there are no Kaluza-Klein modes associated with non-trivial
quantum numbers in $F_2$.  Modular functions which are eigenstates
of the Laplacian cannot be square-integrable, excepting the constant mode.
Indeed, for a square-integrable wave function
obeying $\Box \psi =\lambda \psi $,
one can drop a boundary term in the following integration  by parts
$$
0\leq \int_{F_2} d\mu \ \partial \psi^* \partial \psi
=-\int_{F_2}
d\mu\  \psi^*\Box \psi =-\lambda \int_{F_2}d\mu \  \psi^*\psi \ ,\ \ \
d\mu={dvdx\over v^2}\ ,
$$
which implies $\lambda\leq 0$.
But this contradicts the fact that
the Laplace operator is negative in the fundamental domain (i.e. all
eigenfunctions
have $\lambda\geq 0$).
Thus, scalar modes in $AdS_5\times F_2\times S^4$
will be of the form $\psi =\phi (AdS) Y(S^4) $, with no dependence
on the $F_2$ coordinates.

\subsection{The $AdS_4\times F_3\times S^4$ model}

Let us now consider the $AdS_4\!\times\!H_3/\Gamma\!\times \! S^4$
background.
Here the supergravity solution is given by
\be
ds^2_{11}={R_0^2\over 2z^2}(-dt^2+dx_1^2+dx_2^2+dz^2)+
{R_0^2\over 3v^2}(dx^2+dy^2+dv^2)+{R_0^2\over 4}d\Omega_4^2\ ,
\ee
\be
\hat F_4=6 R_0^3 \Omega_4\ ,
\ee
In ref. \cite{ARVO}, it was considered the case when $H_3/\Gamma$ is  the
hyperbolic dodecahedron space,
which is a compact hyperbolic manifold obtained by identifying opposite
 faces of a dodecahedron after a $3\pi/5$ rotation.
More generally,  we will show below
that any finite volume space of the form $H_3/\Gamma $ does not admit
Killing spinors.
We will mainly be interested in the case when $H_3$ is a non-compact
hyperbolic manifold given by the fundamental region of the Picard group,
which we describe next.

The space $H_3$ finds its natural description in terms of the
quaternionic upper half 3-space (further details can be found in
\cite{terras}). To define it, it is useful to recall
a few known facts about quaternions, which are objects of the form
$$
q=x+i y+j u+k v\ ,
$$
with $x,y,u,v$ real, where
$$
 ij=k=-ji \ ,\ \ jk=i=-kj \ , \ \ ki=j=-ik\ ,\ \ \ \ i^2=j^2=k^2=-1
$$
The norm of a quaternion is given by
$||q||=\sqrt{x^2+y^2+u^2+v^2}\ .
$
Quaternions can be represented in terms of Pauli matrices, so that
$SU(2)$ can be viewed as the unit quaternion.

The quaternionic upper half 3-plane is defined as follows:
\be
H_3=\{ x+i y+k v |x,y,v\in {\bf R}, v>0\}
\ee
Thus the elements of $H_3$ are quaternions with $j$-coordinate equal
to zero. It can be identified with the symmetric space $SL(2,{\bf
C})/SU(2)$.
A matrix $g\in SL(2,{\bf C})$ acts on an element $q$ of the quaternionic
upper half plane
in the following way:
\be
q'\equiv g(q)=(aq+b)(cq+d)^{-1}\ ,\ \ \ \ \ \ g=\pmatrix{ a &b\cr c &d }\
\ee
with $v'=v \ ||cq+d||^{-2}$.
The metric, volume element, and Laplacian on $H_3$
are given by
\ba
&& ds^2 =(dx^2+dy^2+dv^2)v^{-2}\ ,
\\
&& d\mu =v^{-3} dxdydv\ ,
\\
&& \Delta = v^2(\partial^2/\partial x^2+
\partial^2/\partial y^2+\partial^2/\partial v^2)-
v\partial/\partial v\ .
\ea

The Picard group is defined to be
\be
\Gamma=SL(2,{\bf Z}[i])=\{\ g=\pmatrix{ a &b\cr c &d }|a,b,c,d\in
{\bf Z}[i] , \ \det g=1\
\} \ ,
\ee
with
$$
{\bf Z}[i]=\{ x+i y| x,y\in {\bf Z}\}
$$
A fundamental domain for $SL(2,{\bf Z}[i])\setminus H_3  $ in the
quaternionic upper half plane
is then
\be
F_3=\{ x+iy+kv\ |\ |x|\leq 1/2, \ 0\leq y\leq 1/2 ,\ x^2+y^2+v^2\geq 1, \
v>0\}\ .
\ee
The volume of the fundamental domain $F_3$ is \cite{terras}
\be
{\rm Vol}(F_3)=\int_{F_3} d\mu ={|D|^{3/2} \zeta_K (2)\over 4\pi^2}\ ,
\ee
where $\zeta_K(2)$ is the Dedekind zeta function of the number field
${\bf Q}(i)$.

Although non-supersymmetric, the space $AdS_4\times F_3\times S^4$
is an interesting setup for string compactification, in particular,
for the construction of conformal field theories in $2+1$ dimensions (see
sect. 6).

One can show that there is no supersymmetric compactification in
supergravity
involving $F_3$.
Indeed, it is easy to see that no Killing spinor survives after
dividing $H_3$ by $SL(2,{\bf Z}[i])$. The isometry group of $H_3$ is
$SO(1,3)\cong SL(2,{\bf R})\times SL(2,{\bf R})$.
There are two Killing spinors which are Weyl spinors of each $SL(2,{\bf R})$
factor.
The Picard group $SL(2,{\bf Z}[i])$ is essentially given  by
the subgroup $SL(2,{\bf Z})\times SL(2,{\bf Z})$ of $SL(2,{\bf R})\times
SL(2,{\bf R})$.
Therefore it
does not leave invariant any of the
Killing spinors.
More generally, any finite volume example of the form $H_3/\Gamma$
will not have Killing spinors, since the finite-volume condition implies
that
$\Gamma $ must contain elements of both factors in $SL(2,{\bf R})\times
SL(2,{\bf R})$.


\section{Newton's law on $H_3$}

Having described a number of string compactifications containing hyperbolic
spaces,
a question of interest (in view of the recent results of \cite{RS} for
Newton's law in certain non-compact spaces)
is whether such compactifications may be compatible with the observed
Newton's law.
In the case of compactification on a compact space $X$, there exist massive
Kaluza-Klein
modes of the four-dimensional graviton
associated with the eigenvalues of the Laplace operator on $X$.
In the static limit, these modes
contribute to the Newton's law giving, in addition to the
standard $1/r$ behavior of the massless graviton, Yukawa-type
corrections. In particular, these are the expected corrections if the
internal space is a compact hyperbolic manifold of the form $H_r/\Gamma $.
What is the form of the corrections if the internal space is
the non-compact, infinite volume
hyperbolic manifold $H_r$? For non-compact spaces, the spectrum of the
Laplace operator
is continuous. For a continuum spectrum of eigenvalues,
the gravitational potential is given by \cite{KS}
$$
V(r) \sim {1 \over r}\ f(r)\ ,
$$
where
$$
f(r)=\int dq\ e^{-q r} |\Psi_q(0)|^2\ ,
$$
and
$$
-\nabla^2\Psi_q(\rho)=q^2\Psi_q(\rho)\ .
$$
The standard $1/r$
behavior can still be recovered in some non-compact spaces \cite{RS}.

Let us consider for simplicity $M^4 \times H_3$, where $M^4$ is the
four-dimensional
Minkowski space.
Then, in ``spherical" coordinates $(\rho,\theta,\phi)$ for $H_3$ and
metric
\be
ds^2(H_3)=d\rho^2+\sinh^2\rho\left(d\theta^2+\sin^2\theta
d\phi^2\right)\, ,
\ee
the eigenmodes are \cite{CS}
\be
\Psi_{q\ell m}(\rho,\theta,\phi)=R_q^\ell(\rho)Y_{\ell
m}(\theta,\phi)\, ,
\ee
with a  continuous spectrum of eigenvalues $q^2=[1,\infty )$.
The functions $Y_{\ell m}$ are spherical harmonics, and the radial
eigenfunctions are the hyperspherical Bessel functions
\be
R_q^\ell(\rho)={(-1)^{\ell+1}\sinh^{\ell}\rho\over
\left(\prod_{n=0}^{\ell}(n^2+k^2)\right)^{1/2} }\
{d^{\ell+1}\cos(k\rho)\over
d(\cosh\rho)^{\ell+1}}\, , ~~~~k^2=q^2-1\, .
\ee
In the $\ell=0 $ sector (i.e. spherical modes with radial dependence only),
we obtain
\be
f(r)=\int_1^{\infty} dq \ \sqrt{q^2-1}\ e^{-qr}=  {K_1(r)\over r}\, ,
\ee
where $K_1(r)$ is the modified Bessel function of the second kind.
The gravitational potential in four dimensions is then
\be
V(r)\sim {K_1(r)\over r^2} \ .
\ee
At  distances $r$ much smaller than the radius of the hyperboloid (here
taken to be 1
for simplicity), one has $V(r)\sim 1/r^3 $.
At large distances $r\gg 1$, the potential  falls off exponentially,
$V(r)=\sqrt{\pi \over 2 r^5} e^{-r}$. This is expected, since there is
a gap in the spectrum.
As there is no any local minimum of $K_1(r)/r$,  a $1/r$ approximated
behavior cannot be obtained at any scale.

\section{Hyperbolic spaces in type IIA*/IIB* supergravities}

\subsection{$AdS_{2p+1}$ as a $U(1)$ fibration and  untwisting}

In ref. \cite{duff},
a type IIA solution of the form $AdS_5\times CP_2\times S^1$
was obtained by starting with $AdS_5\times S^5$, writing the 5-sphere as a
U(1) bundle over $CP_2$ (Hopf fibration) and performing T-duality
along $S^1$.
Now we make an analogous formal construction of a new type IIA solution
containing a hyperbolic space, by starting with a type IIB solution
describing
$AdS_3$ or $AdS_5$, and
performing a similar operation on the
$AdS $ part of the space.
One can write $AdS_{2n+1}$ as a $U(1)$ bundle over a $2n$-dimensional
base space, and then perform T-duality in the time direction, which untwists the fibration.
As pointed out  in \cite{hull}, T-duality in a time-like direction
does not connect type IIA/IIB theory, but rather relates type IIA to a type IIB* and
type IIB to type IIA*, obtained by reversing the signs of RR kinetic terms.

Let us illustrate this by first considering the case of the $AdS_3\times
S^3\times
T^4$ vacuum of type IIB superstring theory, representing the geometry
of the near-horizon D1-D5 brane system.
 The $AdS_3$ space is the hyperboloid
\be
-X_0^2+X_1^2+X_2^2-X_3^2=-R^2\, ,
\ee
or,
\be
-W\bar{W}+Z\bar{Z}=-1\, \ \ \ \ \ Z=X_1+iX_2\ ,\ \ \
W=X_0+iX_3\ .
\ee
For simplicity, we have set the radius equal to one. The extension
to an arbitrary radius is straightforward. By writing
$$
Z={R\over ({r^2\over R^2}-1)^{1/2}}e^{it}\ , ~~~
W=
{ r\over ({r^2\over R^2}-1)^{1/2}}e^{i(t+\varphi)}\ ,
$$
one obtains the $AdS_3$ metric in the following form
\be
ds^2={dr^2+r^2d\varphi^2\over (1-{r^2\over R^2})^2}- R^2 \left(dt+{r^2\over
R^2-r^2}d\varphi\right)^2
\, . \label{ads3}
\ee
The existence of a $U(1)$ symmetry and
a $U(1)$ fibration over a base $H_2$ is manifest
in these coordinates. The fibration is
trivial and can be turned into a non-trivial one after dividing $H_2$ by a
discrete $\Gamma\subset SL(2,{\bf R})$. The form of the metric
eq.~(\ref{ads3})
makes
also manifest the appearance of closed time-like curves.
These can be avoided by unwrapping the time direction or,
equivalently, by considering the universal covering space of $AdS_3$, which
does not
have any closed time-like curve.
For the purpose of obtaining a new  solution of type II supergravity, we can
formally perform a T-duality transformation along the $S^1$ compactified
time direction in eq.~(\ref{ads3}).
By using the standard T-duality
transformation rules \cite{BHO}, we get the dual solution
\be
d\hat{s}^2={dr^2\over (1-{r^2\over R^2})^2}+{r^2\over (1-{r^2\over R^2})^2}
d\varphi^2-R^2 dt^2\, , ~~~~
B_{t\varphi}=-{r^2\over 1-{r^2\over R^2}} \ , \ \ \ \ e^{2\phi}=g^2={\rm
const.}\ .
\ee
There are also  components of the 2-form  and 4-form field strength tensors
given by
\be
\hat F_2={R^2\ r\over (1-{r^2\over R^2})^2} dr\wedge d\varphi\ ,\ \ \ \ \ \
\hat F_4=2 R^2\ dt\wedge \Omega_3\ .
\ee
Remarkably enough, the metric
\be
d\sigma^2={dr^2\over (1-{r^2\over R^2})^2}+{r^2\over (1-{r^2\over
R^2})^2}d\phi^2\, ,
\ee
is  the metric of  the upper-half plane $H_2$.
Thus, the T-dual of the $AdS_3\times S^3$
vacuum is the space ${\bf R} \times H_2\times S^3$, where we have unwrapped
the
time direction.

Let us now consider the anti de Sitter space $AdS_5$.
In terms of complex coordinates, it is the
hyperboloid:
$$
-W\bar W+Z_1\bar Z_1+Z_2\bar Z_2=-R^2 \ .
$$
Write
$$
W={R\ e^{it}\over (1-{r^2\over R^2})^{1/2}}\ ,\ \ \  Z_1={r \over
(1-{r^2\over R^2})^{1/2}} e^{i(t+{\chi+\varphi\over 2})}\sin{\theta\over 2}
\ ,\  \ \
$$
$$
Z_2={r\over (1-{r^2\over R^2})^{1/2}} e^{i(t+{\chi-\varphi\over
2})}\cos{\theta\over 2}\ .
$$
The metric becomes
\be
ds^2(AdS_5)=d\Sigma_4^2 -R^2(dt-A)^2\ , \ee where
\be
d\Sigma_4^2={dr^2\over (1-{r^2\over R^2})^2}+ {1\over 4}{r^2\over
(1-{r^2\over R^2})}
\left(d\theta^2+\sin^2\theta d\varphi^2\right)+{1\over 4}{r^2\over
(1-{r^2\over R^2})^2}\left(d\chi-\cos\theta d\varphi\right)^2 \, ,
\label{s4}
\ee
\be
A={r^2\over 2 ( R^2- r^2) }\big(d\chi -\cos\theta
d\varphi\big) \ .
\ee
Now we make T-duality in time direction. This
gives a new type IIA* supergravity solution. 
After T-duality, we get
\be
ds^2_{10A}=d\Sigma_4^2 -R^2dt^2 +R^2 d\Omega^2_5\ ,\ \ \ \ \ B_2=
R^2dt\wedge
A\ ,\ \ \ \ \ e^{2\phi}=g^2\ .
\ee
There is also the 4-form field strength $dA_3$ with
components in $\Sigma_4$ implied by the T-duality rules.
Thus
$AdS_5\times S^5$ gets untwisted to a space ${\bf
R}\times\Sigma_4\times S^5$. The space $\Sigma^4$ has self-dual
Weyl tensor and, in fact, it is the constant negative-curvature
partner of $CP_2$, with Ricci tensor
$$
{\cal R}_{\mu\nu}(\Sigma_4 )= -{6\over R^2} g_{\mu\nu}(\Sigma_4 )\ .
$$
Indeed, after $r\to ir$, we find that
$d\Sigma_4^2\to -ds^2(CP_2)$. It is not difficult to see that the space
$\Sigma_4$ is
the coset space $SU(2,1)/SU(2)\times U(1)$. In general, the $AdS_{2n+1}$ can
be
untwisted to $\Sigma_{2n}\times S^1$ where $\Sigma_{2n}$ is
the constant negative-curvature partner of $CP_n$ with metric
obtained from the latter after $r\to ir$ in the Fubini-Study
metric of $CP_n$. $\Sigma_{2n}$ is in fact the coset space
$SU(n,1)/SU(n)\times U(1)$. In the appendix we calculate the 
Euler number of this space.

\subsection{Solutions with fluxes on $H_r$}

A conformal  field theory involving the upper half 3-space $H_3$
can be constructed as a WZW model based on  the coset  $SL(2,{\bf
C})/SU(2)$. This
can be combined with another CFT by direct product.
In particular, it can be combined with the
NS5-brane by summing both conformal sigma models (similar constructions have
been
done in \cite{PRT}). To saturate the central
charge, one needs to add a linear dilaton in the time direction.
The construction leads to an NS-NS 2-form gauge field with imaginary components.
By S-duality, one can convert it into a R-R 2-form with imaginary components,
and thus into a solution of type IIB* theory.
The resulting conformal model is an exact solution of string theory to all
orders in the $\alpha ' $-expansion.

These sigma models can also be constructed directly by brane intersections
(the basic rules for constructing intersecting branes are in
\cite{tseytlin}).
We first describe simpler string solutions
based on
the embedding $H_3$ in ${\bf R}^{10}$ (the extension to other $H_r$ is
straightforward). Let us consider spaces of the form $X_4\times M_6$,
with metric:
\be
ds^2=-dt^2+ t^2 dH_3^2+ds^2_6\  ,
\ \ \ \ \ \ dH_3^2={1\over z^2}(dx^2+dy^2+dz^2)\ ,
\label{rrtt}
\ee
where $ds_6^2$ describes the six-dimensional compact manifold $M_6$.
The hyperbolic 3-manifold
$H_3$ arises
as constant time slices of $X_4$.
The Riemann tensor for $X_4$ vanishes identically. To see this, one
introduces new coordinates
\be
U=t/z\  ,\ \ \ \  X= tx/z\ ,\ \ \  Y= t y/z\ ,\ \  \
V=tz+{t\over z} (x^2+y^2)\ ,
\label{embe}
\ee
so that the metric for $X_4$ takes the form
\be
ds^2_4=-dUdV+dX^2+dY^2\ .
\ee
Thus eq.~(\ref{embe}) provides the embedding of $H_3$ in a four-dimensional
space-time.
If $M_6=T^6$, then the metric (\ref{rrtt}) is an exact solution of string
theory, since the
Riemann tensor identically vanishes. To leading order in $\alpha' $, one can
take any Ricci-flat space $M_6$.
{}From eq. (\ref{embe}) one gets
\be
t^2=UV-X^2-Y^2\, .
\ee
The condition that $t^2\geq 0$ implies that the physical space-time is in
the region
$UV\geq X^2+Y^2$.

 To construct solutions with brane charges, a natural starting point
is the NS5 brane, with flux on $H_3$. It is given by
\be
ds^2_{10}=dx_i^2+ f(t)\big(-dt^2+t^2 dH_3^2\big)\ ,\ \ \ \ \
dH_3^2={1\over z^2}\big(dx^2+dy^2+dz^2)\ ,\ \ \ i=1,...,6
\label{fiv}
\ee
\be
e^{2\phi}=f(t)\ ,\ \ \ \ \ i dB=*df\ ,\ \ \
\ \ \ \ f(t)=1+{R^2\over t^2}\ .
\ee
This is a formal analog of the NS five brane solution, in which
the 3-sphere has been replaced by a hyperbolic space.
Note that the gauge field $dB$ is imaginary.
By S-duality, and redefining the RR 2-form by a factor $i$, 
this is converted into a solution of type IIB* supergravity.

To have finite flux, the space $H_3$ can be replaced by the finite volume
$F_3$ space, representing the fundamental region of the Picard group, as
described previously (or by any coset $H_3/\Gamma $ of finite volume).

By an appropriate rescaling of variables, one finds the
``near horizon" solution:
\be
ds^2_{10}=dx_i^2 -d\tau ^2+ R^2 dH_3^2\ ,\ \ \ \ \ \phi=R^{-1}\tau\ ,\ \ \
\tau=-\ln t.
\label{ner}
\ee
This is a direct product of flat space with linear dilaton and a
$SL(2,{\bf C})/SU(2)$ WZW model.

We now generalize the solution (\ref{fiv}) by considering the intersection
with an NS5 brane
\be
ds^2_{10}=dx_1^2+dx_2^2+f(t)\big(-dt^2+t^2 dH_3^2\big)
+g(r)\big(dr^2+r^2 d\Omega_3^2\big)\ ,
\ee
\be
e^{2\phi}=f(t)g(r)\ ,\ \ \ \ \ i dB= *df(t)+ *dg(r)\ ,\ \ \
g(r)=1+{R_0^2\over r^2}\ .
\ee
Similarly, the NS5-brane can be combined with the solution
(\ref{ner}), giving
\be
ds^2_{10}=-d\tau^2+ dx_1^2+dx_2^2+R^2dH_3^2+
g(r)\big(dr^2+r^2 d\Omega_3^2\big)\
\ee
\be
e^{2\phi}=e^{{2\tau\over R}} g(r)\ .
\ee
One can also add a fundamental string in the direction $x_1$ as follows:
\be
ds^2_{10}=h^{-1}(r)\big[-d\tau^2+ dx_1^2\big]+ dx_2^2+R^2dH_3^2+
g(r)\big(dr^2+r^2 d\Omega_3^2\big)\ ,
\ee
\be
e^{2\phi}=e^{{2\tau\over R}} {g(r)\over h(r)}\ ,
\ \ \ \ \ h(r)=1+{R_1^2\over r^2}\ .
\ee
The near-horizon geometry is given by
\be
ds^2_{10}={r^2\over R_1^2}\big[-d\tau^2+ dx_1^2\big]+R_0^2{dr^2\over r^2}
+ R^2 dH_3^2 + R_0^2 d\Omega_3^2\ ,
\ee
\be
e^{\phi}=e^{{\tau\over R}} {R_0\over R_1}\ .
\ee
This describes a background $AdS_3\times S^3\times H_3\times S^1$,
with a linear dilaton in the time direction.

By U-duality, one can construct
different D-brane solutions with time dependence.
In particular, starting with (\ref{fiv}) and making S and T-dualities,
one obtains solutions with D$p$ brane charge, with $p> 3$:
\be
ds^2_{10B}=f^{-1/2}(t)\big( dx_1^2+...+dx^2_{p+1}\big)
+f^{-1/2}(t)\big[ -dt^2+ t^2dH^2_{8-p}\big]
\ee
\be
e^{2\phi}=f^{(3-p)/2}\ ,\ \ \ \ \ \ \hat F_{8-p}=i\  R^{7-p} \hat H_{8-p}\ ,\ \
\ \ \ \ f=1+{R^{7-p}\over t^{7-p}}\ ,
\ee
where $\hat H_r$ is the volume form of $H_r$.
In the case of $p=3$, one gets near $t=0$ a direct product of de Sitter
space-time and  the hyperboloid $H_5$,
 i.e.
 $dS_5\times H_{5}$ (which can also be obtained
from the $S^5\times AdS_5$ solution by Wick rotation).

\section{Discussion }


The backgrounds described in  sect. 2 can be used  
for
the construction of new
${\cal N}=0$  conformal field theories by holography.
 A version  of the Maldacena dualities  states that string/M theory on
any space of the form
$AdS_{d+1}\times X$ is equivalent to a conformal field theory in
$d$ dimensions. In particular, in the case of the
$AdS_7\!\times\!S^4$ compactification, the boundary theory is a CFT
for  the six-dimensional ${\cal N}=2$ tensor multiplet \cite{Mald}, or a CFT
for the  six-dimensional ${\cal N}=1$ vector multiplet in the
$AdS_7\!\times\!S^4/\Gamma$ background \cite{FKPZ}.
According to this, string (or M) theory compactified on spaces of the form
$AdS_{d+1}\times (H_{r}/\Gamma ) \times S^{n}$
defines a
$d$ dimensional conformal field theory with
$SO(n+1) $ global symmetry.
The correlation functions of the CFT are
 defined by the prescription \cite{gkp,WW}
\be
\langle e^{\int d^dx \ \phi_0(x) {\cal O}(x)}\rangle _{\rm CFT}\equiv 
Z_{\rm string}(\phi ({\rm boundary})=\phi_0(x) ) \ ,
\ee
where $Z_{\rm string}$ is the string theory partition function
computed with boundary values $\phi_0 (x)$ of the string theory fields,
which act as sources of CFT operators.
As usual, in the  classical supergravity approximation,
it can be evaluated as $Z_{\rm string}\cong \exp [-I_{SG}(\phi) ]$,
by solving the equations of motion of $\phi $ in the background
for $AdS_{d+1}\times (H_{r}/\Gamma ) \times S^{n}$,
with the boundary condition $\phi ({\rm boundary})=\phi_0(x)$. 
Consider, in particular, the $d=11$ supergravity solution
$AdS_5\times F_2\times S^4$.
This should be dual to a $d=4$ non-supersymmetric CFT
with $SO(5)$ global symmetry group.
Because the supergravity solution has M5-brane charge,
one may expect the CFT to be related to
the six-dimensional $(2,0)$ or $(1,0)$ conformal field theories.
As explained in sect. 3, eigenfunctions of the Laplace operator on $F_2$
are not square-integrable, barring the constant mode. Thus
Kaluza-Klein modes obeying the Laplace equation are constant  in $F_2$.
In the absence of supersymmetry, finding the field theory
degrees of freedom is nevertheless difficult because the field theory is
expected to be
strongly coupled
in the regime the supergravity description
is valid.

Other interesting cases are the eleven-dimensional supergravity solution
$AdS_4\times F_3\times S^4$, and the type IIB solution
$AdS_3\times F_2\times S^5$. Since the latter has D3 brane charge, it
should be dual to some ${\cal N}=0$ $d=1+1$ conformal
field theory related to ${\cal N}=4$ $d=3+1$ super Yang-Mills theory.
The $AdS_4\times F_3\times S^4$ solution has M5 brane charge.
The dual field theory should be a  $d=2+1$ conformal field
theory
related to the $6d$ $(2,0)$ CFT, with an internal global
symmetry group given by $ SO(5)$.

\bigskip\bigskip

\centerline{\bf Acknowledgements }
We would like to thank A. Tseytlin for useful comments, and
S. Randjbar-Daemi for an important remark. 
A.K. is supported by a $\Gamma.\Gamma$.E.T. grant No. 97/$E\L$/71.
 J.R. would like to thank the support of UBA,
Conicet and Fundaci\' on Antorchas.

\newpage

\noindent
{\bf APPENDIX}
\vspace{.3cm}

In sect. 5 we have seen that the anti de Sitter spacetime $AdS_{2p+1}$
can be represented as a $U(1)$ fibration over a base space 
$\Sigma_{2p}=SU(p,1)/(SU(p)\times U(1))$.  
Here we present  the calculation of the Euler number
of $\Sigma_4$, which  we expect will be of use for future applications.
The space $\Sigma_4$ is a homogeneous space of
the group $SU(2,1)$ and it is diffeomorphic to $SU(2,1)/SU(2)\times U(1)$. Its
metric is given in eq.~(\ref{s4}) and by using the standard $SU(2)$
left-invariant one-forms $\sigma_i,~i=1,2,3, ~~(d\sigma_1=2\sigma_2\wedge
\sigma_3,...)$ on the $S^3$, we may express eq.~(\ref{s4}) as
\be
d\Sigma_4^2={dr^2+r^2\sigma_3^2\over
(1-r^2)^2}+{r^2(\sigma_1^2+\sigma_2^2)\over (1-r^2)}\, ,
\label{ss4}
\ee
where we have put $R=1$. Recalling that the metric on $CP_2$ is
\be
dCP_2^2={dr^2+r^2\sigma_3^2\over
(1+r^2)^2}+{r^2(\sigma_1^2+\sigma_2^2)\over (1+r^2)}\, ,
\label{ss5}
\ee
we see that the above metrics eqs.(\ref{ss4},\ref{ss5}) on
$\Sigma_4=SU(2,1)/SU(2)~\times ~U(1)$ and
$CP_2=SU(3)/SU(2)\times U(1)$ are related by
$r\to ir$. Note that although $CP_2$ is compact, $\Sigma_4$ is
not and it has a boundary at $r=1$.

The Euler number $\chi(\Sigma_4)$ is calculated by the formula
\be
\chi(M)={1\over 32\pi^2}\int_M \epsilon^{abcd}R_{ab}\wedge R_{cd}
-{1\over 32\pi^2}\int_{\partial
M}\epsilon^{abcd}\left(2\theta_{ab} \wedge R_{cd}-{4\over
3}\theta_{ab}\wedge \theta_{ce}\wedge\theta_{ed}\right)\, .
\ee
The first term is the standard bulk contribution whereas
the last term is for the boundary corrections.
A straightforward calculation for $\Sigma_4$ gives
\be
\chi(\Sigma_4)=3\left(1-{1-2r_0\over
(1-r_0^2)^2}\right)-{3r_0^2+2r_0^4\over (1-r_0^2)^2}+{1+2r_0^2\over
(1-r_0^2)^2}=1\, .
\ee
On the other hand, the Euler number of $CP_2$ is $\chi(CP_2)=3$ so
the Euler numbers of $CP_2$ and $\Sigma_2$ differ by two units.
This difference is due to the bold singularity that exists in the
$CP_2$ space and it is absence in the case of $\Sigma_4$. Indeed,
the $CP_2$ metric for $r>>1$ turns out to be
\be
dCP_2^2\sim \r^{-4}\left(dr^2+r^2
\sigma_3^2\right)+\sigma_1^2+\sigma_2^2\, ,
\ee
and the surface $r\to \infty$ is a removable bolt singularity. The
space $\Sigma$ does not have any bolt.
Since each bolt contributes two units to the Euler number, the
Euler number
of $CP_2$ will exceed the corresponding one of $\Sigma_4$ by two,
as indeed happens.


\begin{thebibliography}{99}


\bibitem{DNP}M.~J.~Duff, B.~E.~Nilsson and C.~N.~Pope,
Phys.\ Rept.\ {\bf 130}, 1 (1986).

\bibitem{Mald}J. Maldacena, Adv. Theor. Math. Phys. {\bf 2} (1998) 231,
hep-th/9711200.


\bibitem{horo}G.~T.~Horowitz and A.~Strominger,
Nucl.\ Phys.\ {\bf B360}, 197 (1991).


\bibitem{GT}G.W.~Gibbons and P.K.~Townsend,
Phys. Rev. Lett. {\bf 71} (1993) 3754.

\bibitem{BF}
P.~Breitenlohner and D.~Z.~Freedman,
Annals Phys.\  {\bf 144}, 249 (1982).

\bibitem{FY}Y. Fujii and K. Yamagishi,
 J. Math Phys. {\bf 27} (1986) 979.

\bibitem{LP}H.~Lu, C.~N.~Pope and P. ~Townsend,
Phys.\ Lett.\  {\bf B391}, 39 (1997), hep~-~th/9607164;\\
 H.~Lu, C.~N.~Pope and J.~Rahmfeld,
J.\ Math.\ Phys.\  {\bf 40}, 4518 (1999), hep-th/9805151.

\bibitem{M} J. Milnor, Bull. Math. Soc. {\bf 6} (1982) 9.

\bibitem{ARVO} I.Ya. Aref'eva and I.V. Volovich,
Nucl.Phys. {\bf B274} (1986) 619.

\bibitem{kaloper}N.~Kaloper, J.~March-Russell, G.~D.~Starkman and
M.~Trodden,
``Compact hyperbolic extra dimensions: Branes, Kaluza-Klein modes and
cosmology'', hep-ph/0002001.

\bibitem{kach}
S.~Kachru and E.~Silverstein,
Phys.\ Rev.\ Lett.\  {\bf 80}, 4855 (1998),
hep-th/9802183.

\bibitem{NV}
A.~Lawrence, N.~Nekrasov and C.~Vafa,
Nucl.\ Phys.\  {\bf B533}, 199 (1998),
hep-th/9803015.

\bibitem{cremmer}E.~Cremmer, I.~V.~Lavrinenko, H.~Lu, C.~N.~Pope, K.~S.~Stelle and T.~A.~Tran, ``Euclidean-signature supergravities, dualities and instantons,''Nucl.\ Phys.\ {\bf B534}, 40 (1998), hep-th/9803259.

\bibitem{hull}
C.~M.~Hull,
``Timelike T-duality, de Sitter space, large N gauge theories and  topological field theory,''
JHEP {\bf 9807}, 021 (1998), hep-th/9806146.

\bibitem{hullk}
C.~M.~Hull and R.~R.~Khuri,
``Worldvolume theories, holography, duality and time,''
hep-th/9911082.

\bibitem{FKPZ}
S.~Ferrara, A.~Kehagias, H.~Partouche and A.~Zaffaroni,
Phys.\ Lett.\  {\bf B431}, 42 (1998),
hep-th/9803109; Phys.\ Lett.\  {\bf B431}, 57 (1998)
hep-th/9804006.

\bibitem{DGM}
M.~R.~Douglas and B.~R.~Greene,
Adv.\ Theor.\ Math.\ Phys.\  {\bf 1}, 184 (1998)
hep-th/9707214;
M.~R.~Douglas, B.~R.~Greene and D.~R.~Morrison,
Nucl.\ Phys.\  {\bf B506}, 84 (1997)
hep-th/9704151.





\bibitem{WWW} E. Witten, Nucl.\ Phys.\  {\bf B186}, 412 (1981);\\
L.~Castellani, R.~D'Auria and P.~Fre,
Nucl.\ Phys.\  {\bf B239}, 610 (1984).

\bibitem{CF}
R.~D'Auria, P.~Fre and P.~van Nieuwenhuizen,
Phys.\ Lett.\  {\bf B136}, 347 (1984).

\bibitem{DD}
M.~J.~Duff, B.~E.~Nilsson and C.~N.~Pope,
Phys.\ Lett.\  {\bf B139}, 154 (1984).

\bibitem{PP}
D.~N.~Page and C.~N.~Pope,
Phys.\ Lett.\  {\bf B144}, 346 (1984);
Phys.\ Lett.\  {\bf B145}, 333 (1984).

\bibitem{terras} A.~Terras, ``Harmonics analysis on symmetric spaces and
applications," I, II,
Springer-Verlag 1985.

\bibitem{RS}
L.~Randall and R.~Sundrum,
Phys.\ Rev.\ Lett.\  {\bf 83}, 4690 (1999),
hep-th/9906064.

\bibitem{KS}
A.~Kehagias and K.~Sfetsos,
Phys. Lett. {\bf B427} 39 (2000),  hep-ph/9905417;  A.~Kehagias,
hep-th/9911134.

\bibitem{CS}
N.J.~Cornish and D.N.~Spergel, ``On the eigenmodes of compact hyperbolic
3-manifolds,"
math.DG/9906017.


\bibitem{duff}M.~J.~Duff, H.~Lu and C.~N.~Pope, Nucl.\ Phys.\ {\bf B532},
181 (1998), hep-th/9803061.

\bibitem{BHO}E.~Bergshoeff, C.~Hull and T.~Ort\'\i n,
Nucl.\ Phys.\ {\bf
B451}, 547 (1995), hep-th/9504081.



\bibitem{PRT} G.~Papadopoulos, J.~G.~Russo and A.~A.~Tseytlin,
``Curved branes from string dualities,''
Class.\ Quant.\ Grav.\ {\bf 17}, 1713 (2000), hep-th/9911253.



\bibitem{tseytlin} A.A.~Tseytlin,
Class. Quant. Grav. {\bf 14} (1997) 2085;
hep-th/9702163.





\bibitem{gkp}S.~S.~Gubser, I.~R.~Klebanov and A.~M.~Polyakov,
``Gauge theory correlators from non-critical string theory,''Phys.\ Lett.\ {\bf B428}, 105 (1998),
hep-th/9802109.

\bibitem{WW} E. Witten,
 Adv. Theor. Math. Phys. {\bf 2} (1998) 253,
hep-th/9802150.



\end{thebibliography}
\end{document}